\documentclass[twocolumn,aps,amsmath,amssymb,superscriptaddress,prl]{revtex4-2}
\usepackage{graphicx,bm,amsmath,amssymb,natbib,color}

\usepackage[unicode=true, colorlinks=true, citecolor={blue!80!black}, urlcolor={blue!50!black}, linkcolor = {blue!80!black}]{hyperref}

\usepackage{physics}
\usepackage[normalem]{ulem}
\usepackage{cancel}
\usepackage[dvipsnames]{xcolor}

\makeatletter
\def\clearfmfn{\let\@FMN@list\@empty}    
\makeatother

\begin{document}
\title{Signatures of Interactions in the Andreev Spectrum of Nanowire Josephson Junctions}


\newcommand{\affA}{\affiliation{Departamento de F\'{\i}sica Te\'orica de la Materia Condensada, \mbox{Condensed Matter Physics Center (IFIMAC)} and Instituto Nicol\'as Cabrera, Universidad Aut\'onoma de Madrid, 28049 Madrid, Spain}}
\newcommand{\affB}{\affiliation{Quantronics group, Service de Physique de l'\'Etat Condens\'e \mbox{(CNRS, UMR 3680)}, IRAMIS, CEA-Saclay, Universit\'e Paris-Saclay, 91191 Gif-sur-Yvette, France}}
\newcommand{\affC}{\affiliation{Centro At\'omico Bariloche and Instituto Balseiro, CNEA, CONICET, 8400 San Carlos de Bariloche, R\'io Negro, Argentina}}
\newcommand{\affE}{\affiliation{Center for Quantum Devices, Niels Bohr Institute, University of Copenhagen, Universitetsparken 5, 2100 Copenhagen, Denmark}}
\newcommand{\eqContrib}{\thanks{These authors contributed equally to this work.}}

\author{F. J. \surname{Matute-Ca\~nadas}}
\eqContrib
\affA
\author{C. Metzger}
\eqContrib
\affB
\author{Sunghun Park}
\affA
\author{L. Tosi}
\affC
\author{P. Krogstrup}
\affE
\author{J. Nyg\r{a}rd}
\affE
\author{M. F. Goffman}
\affB
\author{C. Urbina}
\affB
\author{H. Pothier}
\affB
\author{A. \surname{Levy Yeyati}}
\email[Corresponding author: ]{a.l.yeyati@uam.es}
\affA

\begin{abstract}
We performed microwave spectroscopy of an InAs nanowire between superconducting contacts implementing a finite-length, multi-channel Josephson weak link. Certain features in the spectra, such as the splitting by spin-orbit interactions of the transition lines among Andreev states, have been already understood in terms of non-interacting models. However, we identify here additional transitions, which evidence the presence of Coulomb interactions. By combining experimental measurements and model calculations, we reach a qualitative understanding of these very rich Andreev spectra.
\end{abstract}
\maketitle

{\it Introduction.}\textemdash Electronic transport phenomena in nanostructures are typically classified into those which can be understood in terms of non-interacting quasiparticles, such as conductance quantization or universal conductance fluctuations, and those where interactions play a dominant role, such as Coulomb blockade or the Kondo effect \cite{Nazarov-book}. 
In the presence of superconductivity such simplifications are usually fruitful and have allowed the understanding of complex transport properties in the regime of multiple Andreev reflections, when interactions can be neglected \cite{KBT,Shumeiko,Averin,Cuevas,Scheer,Cron} or the interplay between pairing and charging effects in quantum dots with superconducting leads, which can be analyzed in terms of a single level Anderson model \cite{martin-rodero2011,Pillet2013,Medem2019,Kurilovich2021}. As we show in this Letter, microwave spectroscopy experiments on hybrid superconducting weak links (WLs) are, however, challenging such idealized pictures. 

\begin{figure}[t!]
\includegraphics[width=1\columnwidth]{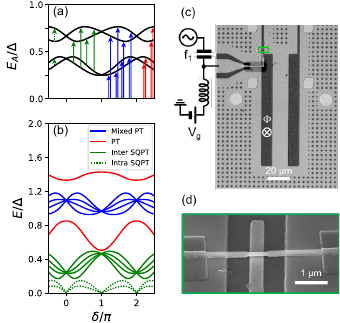}
\caption{(a) Andreev spectrum of a typical finite-length weak link with spin orbit interaction. Colored arrows highlight the possible microwave absorption lines depicted in (b). Transition lines are classified into pair (red) and mixed pair (blue) transitions and intermanifold (solid green) and intramanifold (dashed green) single quasiparticle transitions. (c) Optical image of the measured device, with a phase-biased nanowire (NW) weak link [close wiew in SEM image (d), with  underlying local back gate] placed close to the shorted end of a microwave readout resonator (full view in \cite{supplemental}). Left port connected to the gate is devoted to tune the properties of the NW with a dc voltage $V_g$ and to drive microwave transitions with a tone at frequency $f_1$.}
\label{Fig:Fig1}
\end{figure}

\begin{figure*}[t]
\includegraphics[width=\textwidth]{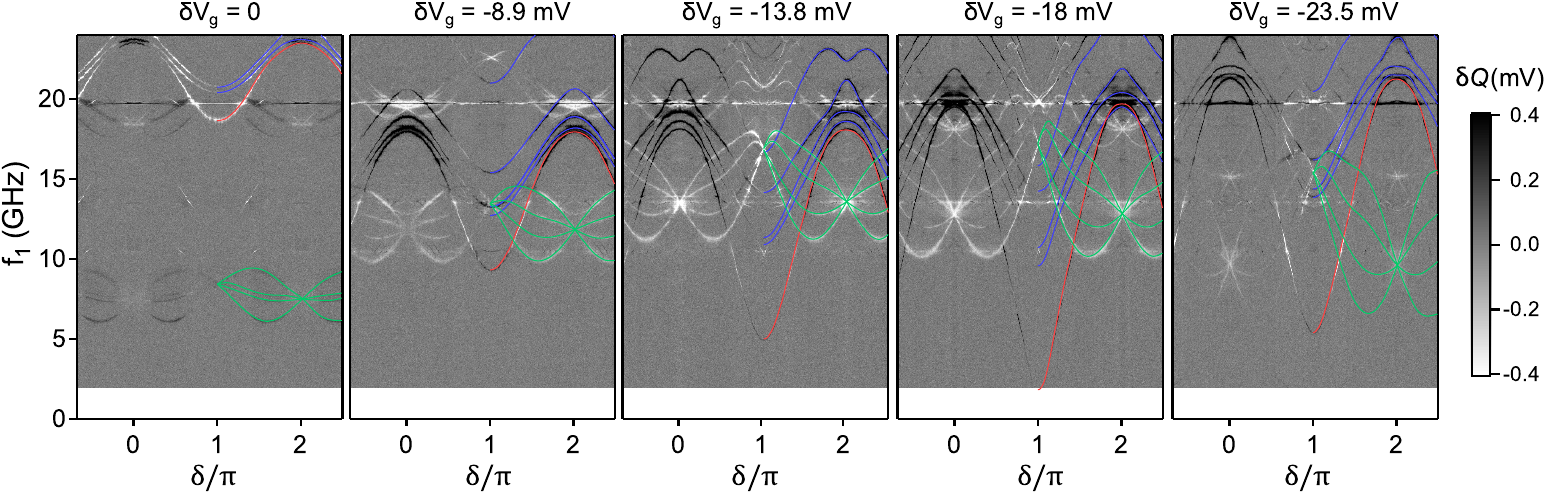}
\caption{Experimental results showing microwave two-tone spectra as a function of phase difference ($\delta$) for a sequence of decreasing gate voltages $V_g=5.563~$V$+\delta V_g$. The colored lines on the right half of the spectra are guides to the eye indicating what we identify as single quasiparticle (green), pair (red) and mixed pair (blue) transition lines. Note that a second group of SQPT is visible around 20~GHz; it likely corresponds to single quasiparticle transitions from the first to the third Andreev doublet (not highlighted here).}
\label{Fig:Fig2}
\end{figure*}

The physics of a Josephson weak link is governed by the spectrum and occupation of its Andreev states. Using circuit-QED techniques \cite{Blais2021} it has been explored and understood in depth for the case of short weak links with a single conduction channel, as in atomic-size contacts \cite{Janvier2015}. Lately, the case of long junctions with several channels is being addressed  owing to the development of high quality hybrid WLs based on semiconducting nanowires \cite{Hays2018,Tosi2019,Hays2020,Metzger2021}. 
Because of spin-orbit interaction, Andreev levels are spin-split for all values of the superconducting phase difference $\delta$ across the link except at the degeneracy points $\delta =0, \pi$, as schematically shown in Fig.~\ref{Fig:Fig1}, where the spectrum is given in the excitation representation. Within a noninteracting model \cite{Tosi2019}, two types of transition lines are expected in the excitation spectrum: on the one hand, single quasiparticle transitions (SQPTs), where trapped quasiparticles in the lower energy Andreev levels are excited into higher ones; on the other hand, pair transitions (PTs), in which two quasiparticles are created on the Andreev levels [see arrows in Fig.~\hyperref[Fig:Fig1]{1(a)}]. The different transition energies are shown in Fig.~\hyperref[Fig:Fig1]{1(b)}, where the green lines correspond to SQPTs while the red and blue ones to the PTs. 
Lines of the red and green type were identified in measured spectra and fitted accurately using such non-interacting models \cite{Tosi2019,Metzger2021}. However, mixed pair transitions in which the two quasiparticles are created in two {\it different} Andreev manifolds [depicted in blue in Fig.~\hyperref[Fig:Fig1]{1(b)}] were never identified.

Here we show that this type of transition is highly sensitive to electron-electron interactions, leading to a modification of their spectral signatures. In contrast, the typical shape of the SQPT lines is not affected by interactions, even though their position with respect to the PTs can change. To discuss the effect of interactions we use different models, ranging from {\it minimal} ones which can be solved exactly to an extended tight-binding model where the effect of interactions is introduced in a perturbative fashion. This extended model allows us to predict spectra that have a close resemblance to those obtained in the measurements.

{\it Experiments.}\textemdash We present here experimental data taken from the sample shown in Figs.~\hyperref[Fig:Fig1]{1(c)} and \hyperref[Fig:Fig1]{1(d)}. A full-shell InAs-Al nanowire 
forms a suspended bridge between the central line and the ground plane of a NbTiN quarter-wavelength coplanar waveguide microwave resonator with resonance frequency $f_r=6.6$~GHz. The Al shell is etched over a $L\sim 550$~nm-long section, defining a Josephson weak link \cite{Goffman2017}. A gate placed under the weak link is dc-biased at voltage $V_g.$ Transitions between Andreev states are driven by microwaves at frequency $f_1$ applied on the gate, and are detected as a change in the reflection coefficient of the resonator. The nanowire shunts the ending part of the resonator, hence defining a superconducting loop that allows phase-biasing the junction: the flux $\Phi$ through this loop imposes the phase difference $\delta=2\pi\Phi/\Phi_0$ across the weak link ($\Phi_0=h/2e$). 
We hardly exceeded the magnetic field of $2~\mu$T required to reach $\delta=2\pi$. 
The coupling between the nanowire and the resonator results from the inductance shared by the loop and the resonator. 

We concentrate here on a series of spectra measured as a function of phase $\delta$ and drive frequency $f_1$, taken successively in a narrow range of gate voltage (Fig.~\ref{Fig:Fig2}). The gray scale represents the change of one quadrature of the measured signal when the drive signal at $f_1$ is applied. Both measurement and drive tones are applied simultaneously, the first one being at the resonator frequency $f_r$. Each pixel corresponds to averaging over 150~ms. 
In the series, certain generic features are observed. There are groups of four lines, such as the ones highlighted in green, which cross at phase $0$ and $\pi$, and are identified as SQPTs. One also finds regular, almost sine-shaped lines, highlighted in red, attributed to PTs. Finally, there are groups of four lines highlighted in blue. They remain grouped  together as the gate voltage is changed, never cross each other, and like pair transitions they have a minimum at $\delta=\pi$. However, they have peculiarities, for instance the ``camel back" shape seen for the topmost blue line at $\delta V_g=-13.8~$mV. As discussed below, these lines can be attributed to mixed pair transitions, in the presence of Coulomb interactions in the nanowire.

{\it Estimations on e-e interactions and their effect.}\textemdash Coulomb interactions in the nanowires are strongly screened
by the nearby metallic electrodes, by free charges in the nanowire and by the substrate.
They can thus be approximated by a contact potential 
\begin{equation}
 \hat{V} = \frac{1}{2}\sum_{\sigma,\sigma'} \int_{\rm WL} d\mathbf{r}d\mathbf{r}' \Psi^{\dagger}_{\sigma}(\mathbf{r})\Psi^{\dagger}_{\sigma'}(\mathbf{r}')u(\mathbf{r}{-}\mathbf{r}')\Psi_{\sigma'}(\mathbf{r}')\Psi_{\sigma}(\mathbf{r})   
 \label{contact-potential}
\end{equation}
where $u(\mathbf{r}{-}\mathbf{r}')=u_0\delta(\mathbf{r}{-}\mathbf{r}')$ is non-zero only for $\mathbf{r},\mathbf{r}'$ in the junction region and $\Psi_{\sigma}(\mathbf{r})$ are the field operators for electrons with spin $\sigma$ in the wire. As we discuss in detail below in the text and in the supplemental material (SM) \cite{supplemental}, we model the wire as a planar quasi-1D geometry. Within this model a rough estimate of the $u_0$ parameter for the experimental situation in Fig. \ref{Fig:Fig1}(c) yields a value of the order of $3~{\rm eVnm^2}$. 

Some insight on the effect of interactions on the energy of Andreev excitations can be obtained by considering the 
analysis of Ref. \cite{Kurland2000} for an isolated mesoscopic grain. In that work it was shown that an interaction as in Eq.~(\ref{contact-potential}) leads to 
an effective exchange interaction $-J \vec{S}^2$, where $\vec{S}$ is the the total spin. The exchange energy is given by $J \sim 2u_0/A$, with $A\approx 0.1 ~\mu {\rm m}^2$ the area where the states are localized, leading to $J \sim 60~\mu$eV (\textit{i.e.} $\sim$10~GHz). As suggested in \cite{Padurariu2012}, such an interaction would lead to a splitting of a group of four mixed pair transitions at $\delta=0$ into a degenerate triplet at lower energy and a singlet state lying roughly $2J$ above. Although this rough analysis fails to explain the full breaking of the degeneracies and the observed complex patterns, one can clearly observe in Fig.~\ref{Fig:Fig2} a tendency of the lines highlighted in blue to group into three lower lines and one higher line, which is reminiscent of a singlet-triplet splitting.    

{\it Tight binding model.}\textemdash A rather simple multi-channel tight-binding (TB) model describing some of the observed features in the Andreev levels absorption spectrum was introduced in Refs.~\cite{Metzger2021,Hays2021}. It corresponds to a discretized version of the model in Ref. \cite{Tosi2019} and is given by
\begin{align}
H_0 &= \sum_{i,\tau,\sigma} (\epsilon_{i,\tau} - \mu)c^{\dagger}_{i,\tau,\sigma}c_{i,\tau,\sigma}+ t_x c^{\dagger}_{i,\tau,\sigma} c_{i+1,\tau,\sigma}\nonumber\\
&+ \sigma \alpha_x c^{\dagger}_{i,\tau,\sigma} c_{i+1,\tau,\bar{\sigma}} + \sum_{i,\tau} \Delta_{i} c_{i,\tau,\downarrow} c_{i,\tau,\uparrow} \nonumber\\
& + \sum_{i,\tau,\sigma} t_y c^{\dagger}_{i,\tau,\sigma}c_{i,\tau+1,\sigma} + i\alpha_y c^{\dagger}_{i,\tau,\sigma}c_{i,\tau+1,\bar{\sigma}} + \mbox{H.c.} \;,
\label{extended-TB-model}
\end{align} 
where $c^{\dagger}_{i,\tau,\sigma}$ creates an electron on longitudinal site $i$, transverse site $\tau$ and with spin $\sigma$; $\epsilon_{i,\tau,\sigma}$ denotes the onsite potential, $t_{x,y}$ and $\alpha_{x,y}$ are spin-conserving and spin-flip hopping amplitudes in the longitudinal and transverse direction respectively, and $\Delta_i$ is the pairing amplitude which we choose to be zero for the sites describing the wire and $\Delta e^{\pm i\delta/2}$ for the left and right superconducting electrodes, respectively.
This model can be adapted to include the effect of interactions in the central normal region $N$ by adding a Hubbard-like term $H_{\rm int} = \sum_{i \in N,\tau} U_i n_{i,\tau,\uparrow} n_{i,\tau,\downarrow}$ while assuming perfect screening in the superconducting regions. Here, $n_{i,\tau,\sigma}=c^{\dagger}_{i,\tau,\sigma}c_{i,\tau,\sigma}$. Setting a given value for the lattice spacings in the $x,y$ directions, $a_{x,y}$, one can get estimates for all the model parameters appropriated for InAs wires coupled to Al leads  by discretizing the continuous model \cite{supplemental}. 
Additionally, the $U_i$ value, taken as a constant $U$ in the normal region, is related to the above $u_0$ estimate by $U \sim u_0/(a_x a_y)$. One can also define an effective charging energy of the normal region when disconnected from the leads $E^{\rm eff}_c {=}u_0/A_N,$ where $A_N$ denotes its area.

This model cannot be solved exactly in the presence of interactions. However, we can get insight into their effect on the subgap states by considering the infinite gap limit $(\Delta_i \rightarrow \infty)$ \cite{Meng2009} and restricting the normal region to four sites only. Such a four-site model is a minimal one that can account for the multichannel character and the finite length of the junction  while being amenable to exact diagonalization including the Hubbard terms. The effect of the superconducting pairing in the leads projected into the central four sites [denoted by $\alpha=L,R$ (left,right) and $\tau=\pm$ (top, bottom)] leads to the following effective pairing model \cite{supplemental}

\begin{figure}
\includegraphics[width=1\columnwidth]{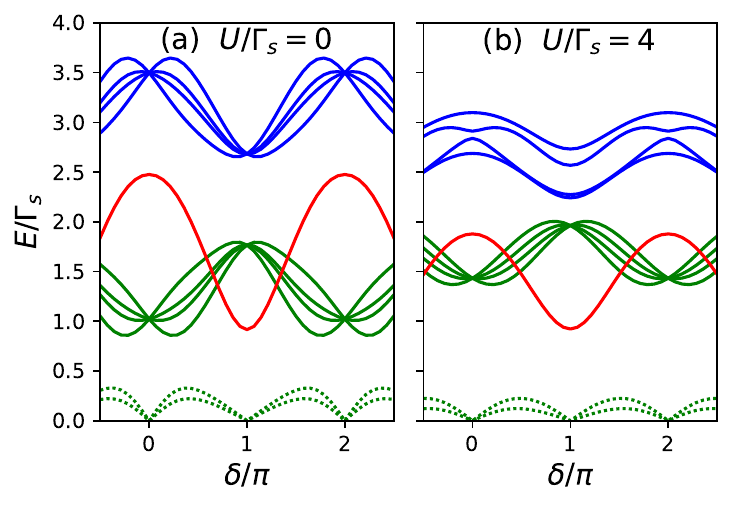}
\caption{Transition lines within the four-site model without (a) and with (b) the effect of Coulomb interactions. Within this model effective singlet and triplet pairing, characterized by parameters $\Gamma_s$ and $\Gamma_t$, arise by assuming $\Delta \rightarrow \infty$ in the leads.}
\label{Fig:Fig3}
\end{figure}

\begin{eqnarray}
H_{\rm pairing} &=& \sum_{\alpha,\tau=\pm} \Gamma_{s,\alpha} c^{\dagger}_{\alpha,\tau,\uparrow} c^{\dagger}_{\alpha,\tau,\downarrow} + \\
&& i\sum_{\alpha} \Gamma_{t,\alpha} \left(c^{\dagger}_{\alpha,+,\uparrow} c^{\dagger}_{\alpha,-,\uparrow} - c^{\dagger}_{\alpha,+,\downarrow}c^{\dagger}_{\alpha,-,\downarrow}\right) + \mbox{H.c.} \nonumber
\end{eqnarray}
where $\Gamma_{s,\alpha}$ and $\Gamma_{t,\alpha}$ are effective singlet and triplet pairing amplitudes for the $\alpha{=}L,R$ sites arising from the combination of s-wave pairing and spin-orbit interactions in the multi-channel leads.
We do not expect that the scaling used to determine the parameters in Eq.~(\ref{extended-TB-model}) would hold for such a minimal system. However, setting reasonable parameters (e.g.\ \(\epsilon_{i,\tau}/2{=}\Gamma_s{=}{-}t_x{=}{-}t_y\) and \(\Gamma_{t}{=}\alpha_x{=}\alpha_y{=}0.8\Gamma_s\)), we get the results shown in Fig.~\ref{Fig:Fig3} for the effect of interactions on the transition lines which have some resemblance with the experimental observations. 

\begin{figure*}[ht!]
\includegraphics[width=0.98\textwidth]{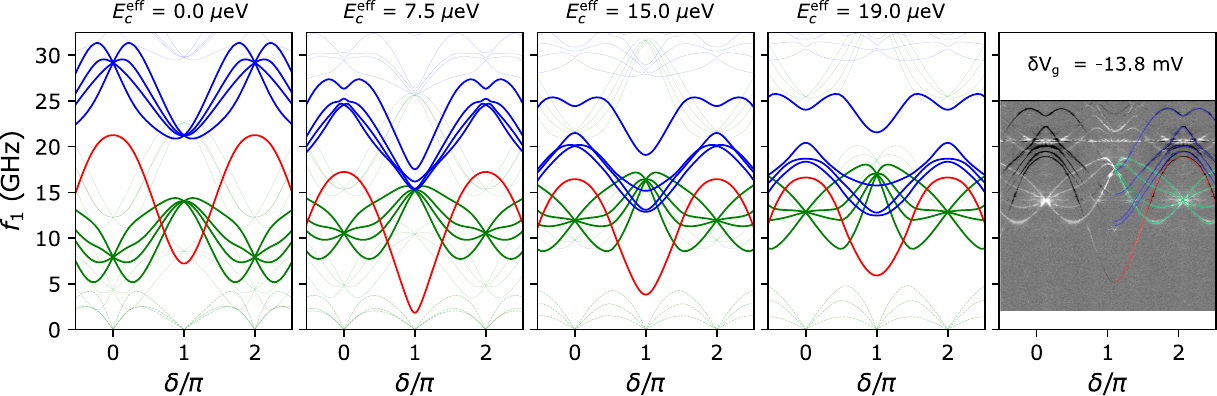}
\caption{Evolution with the weak link effective charging energy $E_c^{\rm eff}$ of the spectral lines as a function of phase difference as obtained from the extended TB model with 31 sites in the $x$-direction (11 in the normal region) and 3 transverse chains, describing a junction with length $\sim 550$~nm and width $\sim 200$~nm (detailed parameters in \cite{supplemental}). Full lines correspond to the main inter SQPT (green), lowest PT (red) and mixed PTs (blue). The faint lines correspond to secondary transitions (\textit{i.e.} from the first to the third or from the second to the third manifolds, intra-manifold and higher PTs). Excitations up to $N_{\rm pr}=12$ are included in the effective interacting Hamiltonians. Rightmost panel are the data of the central panel of Fig.~\ref{Fig:Fig2}.}
\label{Fig:Fig4}
\end{figure*}

As expected, interactions lift the degeneracy of the mixed pair transitions (blue lines in Fig.~\ref{Fig:Fig3}) at $\delta=0$ and $\pi$. Moreover, in contrast to the simple argument based on the emergent exchange interaction, which splits the transition lines into triplet and singlet \cite{Padurariu2012}, there is a complete splitting of the four lines. This is due to the presence of a significant spin-orbit interaction which breaks spin symmetry.

On the other hand, the intermanifold single quasiparticle transition lines (solid green lines in Fig.~\ref{Fig:Fig3}) do not split at $\delta=0,\pi$ but are rather shifted to higher energy. These crossings are protected by time reversal symmetry which leads to a Kramers degeneracy \cite{Kramers} for odd states even in the presence of interactions. The shift to higher energy can be understood as a consequence of level repulsion between the lower and upper Andreev manifold when coupled through the Coulomb interaction. In contrast, no Kramers degeneracy is granted for even parity excitations, which explains the splitting of the mixed transitions.  It should be noticed that the ground state parity remains even within this interaction range (no $0{-}\pi$ transition). An analysis of the phase diagram for this model is given in the SM~\cite{supplemental}.

To go beyond this four-site model we use the eigenstates of Eq.~(\ref{extended-TB-model}) calculated for the case of three chains in the $y-$ and multiple sites in the $x-$ directions to write the interaction Hamiltonian \(H_{\rm int}\) in terms of the Bogoliubov operators $\gamma_n$. This is performed through the inverse Bogoliubov transformation
$c_{i,\tau,\sigma}=\sum_{n\geq1} u^n_{i,\tau,\sigma} \gamma_n + v^{n*}_{i,\tau,\sigma}\gamma^{\dagger}_n$,
where $n{\geq}1$ refers to states with positive energy and \(u^n_{i,\tau,\sigma}\) (\(v^n_{i,\tau,\sigma}\)) are the electron (hole) -like coefficients of the noninteracting wavefunctions. 

Assuming weak interactions, we may project \(H_{\rm int}\) to the subspace of states with zero (\(\ket{GS}\)), one (\(\gamma^{\dagger}_{n}\ket{GS}\)), and two (\(\gamma^{\dagger}_{n}\gamma^{\dagger}_{m}\ket{GS}\)) quasiparticles on the $N_{\rm pr}$ lowest energy levels (i.e. \(n,m{\le}N_{\rm pr}\)). Because of parity conservation we end up with effective Hamiltonians in the even and odd sectors that can be diagonalized exactly. An analysis of the range of validity of this approximation is given in the SM \cite{supplemental}. Using such a procedure, we searched for a set of parameters that reproduce at best the central spectrum of Fig.~\ref{Fig:Fig2}, in which the full dispersion of the lines highlighted in blue is visible. The result is shown next to the data in Fig.~\ref{Fig:Fig4}. Most features of the experiment, both for the relative frequencies and for their shape, are essentially reproduced. In particular, the camel back dispersion of the upper mixed pair line around phase 0, absent in a non-interacting model, is captured. It should be mentioned, however, that these spectra are extremely sensitive to microscopic details in the potential profile, which are completely unknown for an actual experimental realization.
In the other panels of Fig.~\ref{Fig:Fig4}, we show how the spectrum evolves when changing only the Coulomb interaction strength.
As in the case of the four-site model, the most remarkable effect of interactions is to lift the degeneracies of the mixed pair transition lines at $\delta=0,\pi$  (blue lines in Fig.~\ref{Fig:Fig4}) and to shift the intermanifold single quasiparticle lines (green lines in Fig.~\ref{Fig:Fig4}) to higher frequency without breaking their characteristic shape.  

It is worth stressing that the multichannel character of realistic nanowire weak links is essential to understand the shape of the pair transition lines measured experimentally. Indeed, the set of parameters which gives the closest resemblance with the data corresponds to a case with two open channels such that the phase curvatures of the two lowest Andreev manifolds have the same sign \cite{supplemental}. Such a property cannot be obtained with a single channel model, which gives manifolds with alternating curvatures \cite{Bagwell1992}.
Finally, notice that here, as for the case of the four-site model in Fig.~\ref{Fig:Fig3}, the interaction strength, although sufficiently large to produce sizeable effects in the transition lines, is still too weak to produce a transition into the $\pi$ phase as has been observed in the case of quantum dots coupled to superconducting leads \cite{Cleuziou2006,Jorgensen2007,Lee2014}. 

{\it Conclusions.}\textemdash  We have shown that a complete description of the microwave spectrum of semiconducting nanowire Josephson junctions must include Coulomb interactions in addition to spin-orbit multi-channel coupling. This is more clearly evidenced by the splitting of mixed pair transitions around 0 and $\pi$ phase difference. Despite the strength of the required interactions not being enough to drive these systems into a $\pi$ phase, as is usually the case in the quantum dot regime, the high sensitivity of circuit-QED techniques allows us to reveal their presence in actual devices.\\ 

All raw data in the publication are available at \cite{zenodo}.\\ 

Technical support from P. S\'enat is gratefully acknowledged. We thank P. Orfila and S. Delprat for nanofabrication support, and our colleagues from the Quantronics group and C. Strunk for useful discussions. We thank Will Oliver and Lincoln Laboratories for providing us with the TWPA used in the experiment, and  E. Flurin for suggesting the bridge sample geometry. 
We thank V. Fatemi, M. H. Devoret, P. D. Kurilovitch and L. I. Glazman for sharing with us their manuscript \cite{Fatemi2021}, for useful comments on ours, and for fruitful discussions. We also thank M. Houzet and J. Paaske for their comments.
This work has been supported by ANR contract JETS, by FET-Open contract AndQC, the Danish National Research Foundation, by the Renatech network, by the Spanish AEI through Grant No.~PID2020-117671GB-I00 
and through the ``Mar\'{\i}a de Maeztu'' Programme for Units of Excellence in R\&D (CEX2018-000805-M).
C. Metzger was supported by Region Ile-de-France in the framework of DIM SIRTEQ. F.J.M. acknowledges support from the Spanish Ministry of Universities (FPU20/01871).\\

F.J.M. and C.M. contributed equally to this work.\\

\noindent {\it Note added.}\textemdash Recently, we became aware of a related work by V. Fatemi \textit{et al.} \cite{Fatemi2021} also pointing out the relevance of interactions in nanowire junctions.

\onecolumngrid
\newpage

\begin{center}
\large\textbf{Supplementary Information \\Signatures of Interactions in the Andreev Spectrum of Nanowire Josephson Junctions}  
\end{center}
\setcounter{page}{1}

\maketitle
\nopagebreak[0]

\renewcommand{\thepage}{S\arabic{page}}
\renewcommand{\thesection}{S\arabic{section}} 

\begin{center}
\textbf{CONTENTS}    
\end{center}

\noindent
\hyperref[sec:S1]{\color{black}S1. Estimates for the $u_0$ parameter} \hfill \pageref{sec:S1} \\

\noindent
\hyperref[sec:S2]{\color{black}S2. Details on the extended tight-binding model} \hfill \pageref{sec:S2} \\

\noindent
\hyperref[sec:S3]{\color{black}S3. Details on the four-sites model and phase diagram} \hfill \pageref{sec:S3} \\

\noindent
\hyperref[sec:S4]{\color{black}S4. Evolution of the transition lines} \hfill \pageref{sec:S4} \\

\noindent
\hyperref[sec:S5]{\color{black}S5. Experimental details} \hfill \pageref{sec:S5} \\


\section{S1. Estimates for the $u_0$ parameter}
\label{sec:S1}
The typical junction dimensions (length $\sim 500$ nm and diameter $W \sim 140$ nm) and the fact that a few conduction channels might be contributing to transport suggest that a 3D screening model should be appropriate. Within a Thomas-Fermi (TF) approximation we have  
\(u(\vec{r}) \sim e^{-|\vec{r}|/\lambda_{\rm TF}}/|\vec{r}|\), where $\lambda_{\rm TF}$ is the screening length and thus
\begin{align*}
u^{3D}_0 = \frac{e^2 }{4\pi \epsilon_0\epsilon_r} 4\pi \int_{0}^{\infty}
dr r^2 \frac{e^{-r/\lambda_{\rm TF}}}{r} =
\frac{e^2 }{4\pi \epsilon_0\epsilon_r} 4\pi \lambda_{\rm TF}^2
\end{align*}

On the other hand, the TF screening length can be estimated as 
\begin{equation*}
\lambda_{\rm TF}^2= \epsilon_0 \epsilon_r/(e^2\rho) = \frac{a_B}{8}\frac{m_e}{m^*}\epsilon_r \lambda_F,
\end{equation*}
where $\rho = (2 m^*/\hbar^2)^{3/2} \sqrt{E_F}/(2 \pi^2)$ is the 3D density of states with the effective mass $m^* \sim 0.023\, m_e$ and the Fermi energy $E_F$, $a_B= 4\pi\epsilon_0\hbar^2/(m_e e^2) \sim 0.05\, \text{nm}$ is the Bohr radius, $\lambda_F$ is the Fermi wavelength and $\epsilon_r \sim 15$ is the semiconductor dielectric constant. As the data suggest, $\lambda_F$ should correspond to a situation where a second subband starts to be populated, i.e.
$\lambda_F \sim W \sim 140\, \text{nm}$ one gets $\lambda_{\rm TF}\sim 24 \, \text{nm}$. Thus, as $\lambda_{\rm TF} \ll W$ 
a 3D model is justified. We can further get the 2D $u_0$ by $u_0=u^{3D}_0/W\sim \pi\hbar^2/(2m^*)$, obtaining values of $u_0\sim 5~\textrm{eV nm}^2$. 
 Similar values were estimated in Ref. \cite{Stanescu2014}. In spite of what this simple derivation might suggest, in experiments we do not expect that the $u_0$ parameter, and thus the effective charging energy, could be directly controlled by the gate voltage.
 
\section{S2. Details on the extended tight-binding model}
\label{sec:S2}

\begin{figure}[t]
    \centering
    \includegraphics[width=1\linewidth]{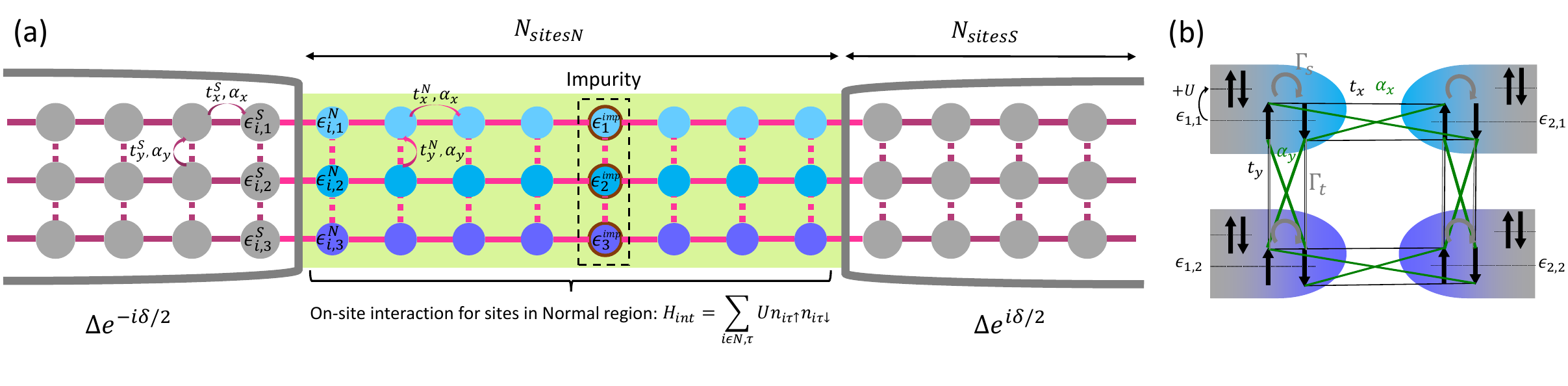}
    \caption{Sketch of the extended (a) and four-sites (b) tight-binding models for the nanowire junction. In the four-sites sketch, the grey shading represents the effect of the superconducting leads projected into each site. This is achieved through the effective singlet and triplet pairings \(\Gamma_s\) and \(\Gamma_t\) (grey arrows and lines) between electrons (thick, black arrows), which have onsite energy \(\epsilon_{i,\tau}\). Other lines depict the spin conserving (thin black) and spin flipping (thin green) hoppings. Finally, the interaction is represented with the gain in energy \({+}U\) when a site is occupied with two electrons.}
    \label{fig:sketch}
\end{figure}

\begin{figure}[h!]
    \centering
    \includegraphics[width=1\linewidth]{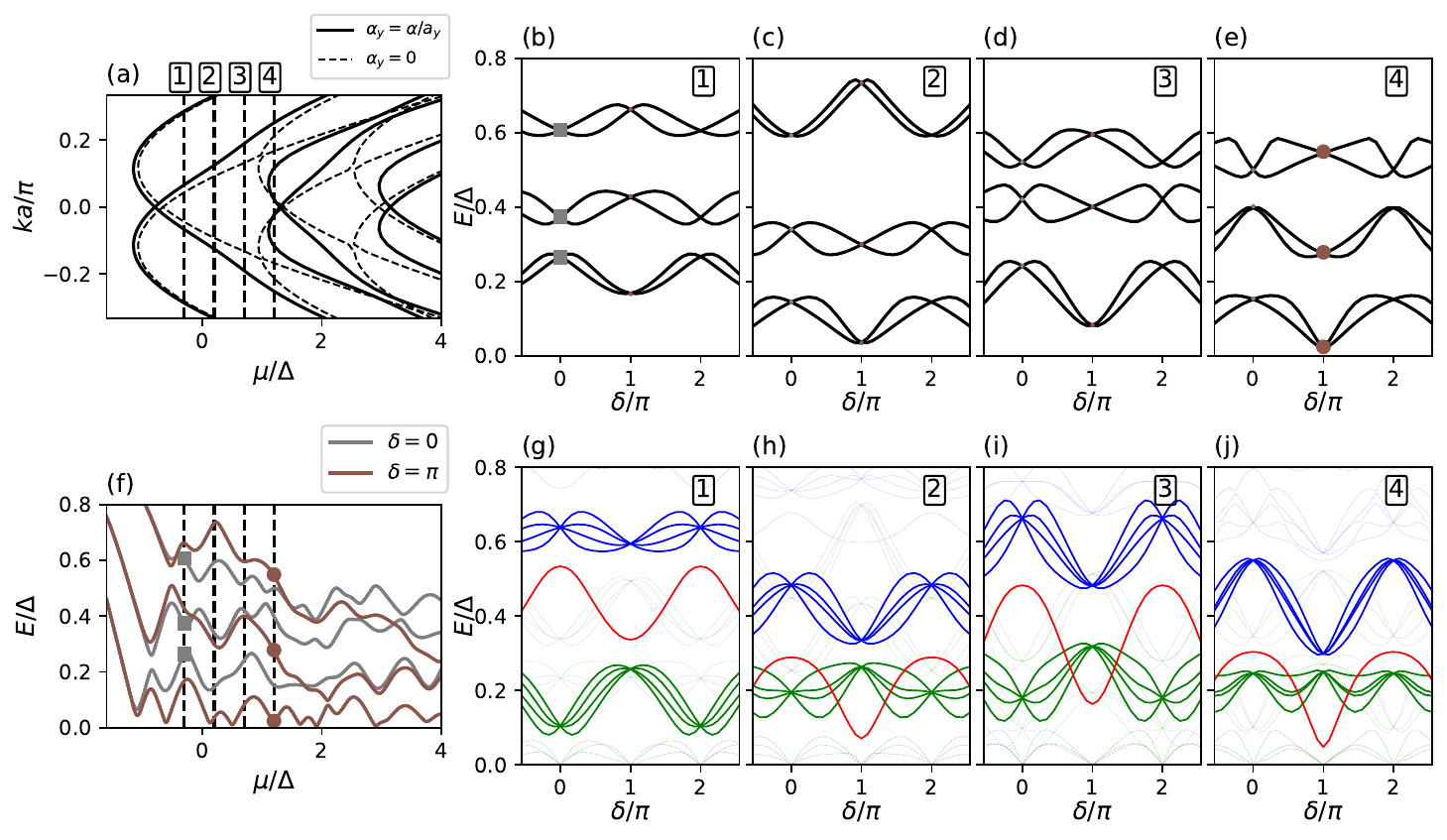}
    \caption{Behaviour of ABSs energies and transitions with $\mu$, obtained from the extended TB model without interaction and the parameters used in Fig. 4 of the main text (specified in the caption of Fig. \ref{fig:NprextendedTB}). (a) Bulk bands of the normal region. Dashed lines depict the bands when setting the transverse SO coupling $\alpha_y$ to $0$ and illustrate that its effect is to hybridize the channels. 
    Vertical dashed lines correspond to 4 different values of $\mu$ for which the energies of the three lowest Andreev manifolds are shown in (b-e), with the corresponding transitions in (g-j), highlighting odd transitions between the lowest two manifolds, the lowest pair transition, and the mixed transitions involving both manifolds 1 and 2 (colour code as in the main text). Cases 1 and 4 correspond to the $\mu$ being placed, respectively, in the 1st and the 2nd subbands, showing a change of the second Andreev manifold curvature around $\delta{=}\pi$ from negative to positive. A more detailed evolution of Andreev energies at $\delta{=}0,\pi$ is shown in (f), where the markers indicate the corresponding points in (b-e). The crossing between the energies of the second manifold with phase difference $0$ and $\pi$ roughly indicates where the curvature of the manifold starts changing.}
    \label{fig:evo_mu}
\end{figure}

\begin{figure}[h!]
   \centering
    \includegraphics[width=1\linewidth]{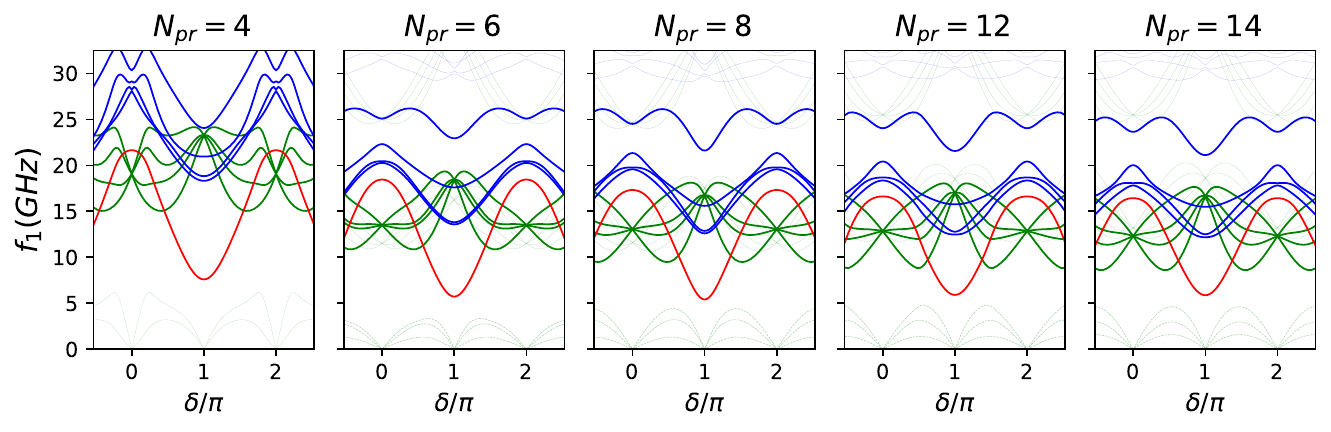}
    \caption{Transitions from the extended TB model increasing \(N_{pr}\) to include progressively higher states. Parameters are the same as in Fig. 4 of the main text: $N_{\rm sitesN}{=}11$, $N_{\rm sitesS}{=}10$, $\mu{=}0.14~$meV (in the N region), $\Delta{=}0.2~$meV,  $L{=}550~$nm, $a_x{=}L/N_{\rm sitesN}$, $a_y{=}100~$nm,   $\alpha/2{=}11~$meV.nm${=}a_x\alpha_x{=}a_y\alpha_y$, \(t_0{=}\hbar^2/2m^*\) (\(m^*{=}0.023m_e\)), $(\epsilon_{1N},\epsilon_{2N},\epsilon_{3N}){=}(1.2,1.1,0.8){\cdot}2t_0/a_x^2$,  $\epsilon_{1S}{=}\epsilon_{2S}{=}\epsilon_{3S}{=}2t_0/a_x^2{-}\Delta$, $(t_{xN},t_{xS}){=}({-}0.85,{-}1){\cdot}t_0/a_x^2$, $t_{yN}{=}t_{yS}{=}{-}t_0/a_y^2$. Impurity position: site $3$ of N region, $(\epsilon_{\rm imp1}, \epsilon_{\rm imp2}, \epsilon_{\rm imp3}){=}(0.6,0.75,0.75){\cdot}2t_0/a^2$.   \label{fig:NprextendedTB} }
\end{figure}

The nanowire transverse channels hybridize in presence of SO coupling. This fact is key to the splitting of the Andreev bound states (ABSs), since the states on each channel acquire a spin texture with different Fermi velocities for a given chemical potential \(\mu\) \cite{Park2017,Tosi2019}. That is why a multi-chain tight-binding model must be considered. 

The parameters for the extended tight-binding model (Eq. (2) in the main text) correspond to a discretization of a continuous Hamiltonian, i.e.
\[ t_{x,y} \sim \frac{\hbar^2}{m^* a_{x,y}^2}; \; \alpha_{x,y} \sim \frac{\alpha}{2a_{x,y}} ,\]
where $a_{x,y}$ are the lattice spacing in the $x,y$ direction and $\alpha \sim 15-30$meVnm is the spin-orbit coupling constant appropriate for InAs. 
We also consider the effect of impurities represented by a different site energy $\epsilon^{\rm imp}_{\tau}$ at certain positions (as sketched in Fig. \ref{fig:sketch}a). These are used as fitting variables to obtain spectra in reasonable agreement with the experimental results and are specified in the caption of Fig. \ref{fig:NprextendedTB}.

As described in the main text, reproducing some relevant features of the experimental lines requires the presence of two open channels. Within our extended tight-binding model this situation is well described when including 3 transverse sites, providing up to 3 hybridized channels. In Fig.~\ref{fig:evo_mu} we show the evolution of the non-interacting ABSs and the associated transition lines as a function of the chemical potential $\mu$, which controls the doping level in the normal region. For reference, we show on the panel (a) the subbands corresponding to an infinite wire where the chemical potential positions used in the other panels are indicated.  As can be observed, only when \(\mu\) approaches the bottom of the second subband, the curvature of the second Andreev manifold appears in phase with that of the first one.  

The eigenstates $\Phi_n$ of the non-interacting model are calculated by diagonalizing the corresponding Bogoliubov-de Gennes Hamiltonian
\[ H_{BdG} \Phi_n {=} E_n \Phi_n,\;\;\;  H_0{=}\frac{1}{2}\hat{\Psi}^{\dagger}H_{BdG}\hat{\Psi},\] 
where \(\hat{\Psi}{=}(\hat{c}_{1,1},\hat{c}_{1,2},\hat{c}_{2,1},\hat{c}_{2,2},...)^T\), \(\hat{c}_{i,\tau}{=}(c_{i\tau\uparrow},c_{i\tau\downarrow},c^\dagger_{i\tau\downarrow},-c^\dagger_{i\tau\uparrow})^T\) and \((\Phi_n)_{i\tau}{=}(u^{n}_{i\tau\uparrow},u^{n}_{i\tau\downarrow},v^{n}_{i\tau\downarrow},-v^{n}_{i\tau\uparrow})^T\). The quasiparticle (QP) operators that diagonalize \(H_0\) are related to the eigenvectors by \(\gamma_n{=}\Phi^\dagger_n \hat{\Psi} \leftrightarrow \hat{\Psi}{=}\sum_{n} \Phi_n \gamma_n\), and the electron-hole symmetry implicit in the BdG formalism, that relates states with opposite energy (\(\gamma^\dagger_n{=}\gamma_{{-}n}\), \(E_{-n}{=}{-}E_n\)), allows to write it in terms of the quasiparticle operators of states with positive energy \(H_0{=}E_{GS}+\sum_{n{\ge}1} E_n \gamma^\dagger_{n} \gamma_n\), where \(E_{GS}{=}1/2\sum_{n{\le}-1}E_{n}\) is the energy of the ground state (GS), in which all states with negative energy are occupied. Thus, quasiparticle excitations over the GS of e.g. 1 and 2 QPs are represented by \(\gamma^{\dagger}_{n}\ket{GS}\) and  \(\gamma^{\dagger}_{n}\gamma^{\dagger}_{m}\ket{GS}\) (\(n,m{\ge}1\)), satisfying \(\gamma_{n}\ket{GS}{=}0\).

As described in the main text, the interaction is introduced by projecting \(H_{int}\) into the many-body states with zero (GS), one and two QP excitations of lowest energy (\(n,m{\le}N_{pr}\)). This requires the calculation of cumbersome expectation values such as \(\bra{GS} \gamma_{i_2}\gamma_{i_1} \gamma^{(\dagger)}_{n_1}\gamma^{(\dagger)}_{n_2}\gamma^{(\dagger)}_{n_3}\gamma^{(\dagger)}_{n_4}\gamma^\dagger_{j_1}\gamma^\dagger_{j_2}\ket{GS}\), which have been computed using the \textit{QuantumAlgebra.jl} package \cite{QuantumAlgebra.jl}. The convergence of the results with the number of states \(N_{pr}\) where the interaction is projected is shown in Fig. \ref{fig:NprextendedTB}, which displays a stabilization for \(N_{pr}>6\) for the interaction strength which gives the best agreement with experiment $(E^{\rm Eff}_c{=}19\mu$eV). For increasing interaction strengths the convergence with $N_{pr}$ becomes slower, which implies a larger mixing with continuum states. As these states are poorly described in our finite size model we expect that results obtained using this method for $U \gg \Delta$ would be less reliable.

\section{S3. Details on the four-sites model and phase diagram}
\label{sec:S3}

A quantum dot (QD) between superconducting leads is a typical nanostructure where the effect of interactions plays a dominant role, and can be typically analyzed in terms of an Anderson model where a single level with Hubbard like interaction is connected to the leads. This model is, however, not able to describe the experimental situation found in Refs. \cite{Tosi2019,Metzger2021} where at least two ABSs associated to the finite length of the junction appear. In addition, the coupling of transverse modes due to spin-orbit interactions is necessary to explain the splitting of SQPT as commented in the previous section. From this reasoning we conclude that a minimal model should include 2 sites both in the longitudinal and transverse directions. On the other hand, to include the superconducting leads in a simplified manner one can take the limit $\Delta_i \rightarrow \infty$ as suggested in several works on the superconducting Anderson model \cite{Vecino2003,Meng2009}. While in the single level model such limit leads to an induced local singlet pairing in the dot, for the case of the multichannel spin-orbit coupled leads we can expect both local singlet and non-local triplet pairings to be induced on the central region. As mentioned in the main text, we indicate by $\Gamma_s$ and $\Gamma_t$ the corresponding effective pairing amplitudes. 
Obtaining their expressions in terms of the bare model parameters would require the calculation of the leads boundary Green functions \cite{Alvarado2020} in the $\Delta_i\rightarrow \infty$ limit. While this calculation could be affordable using the techniques of Ref. \cite{Alvarado2020}, in the present work we just consider $\Gamma_{s,t}$ as tunable effective parameters. The good qualitative agreement with the experimental results and with the calculations from the extended TB model gives us indication that this effective pairing model corresponds to a reasonable approximation.

\begin{figure}[t]
    \centering
    \includegraphics[width=1\linewidth]{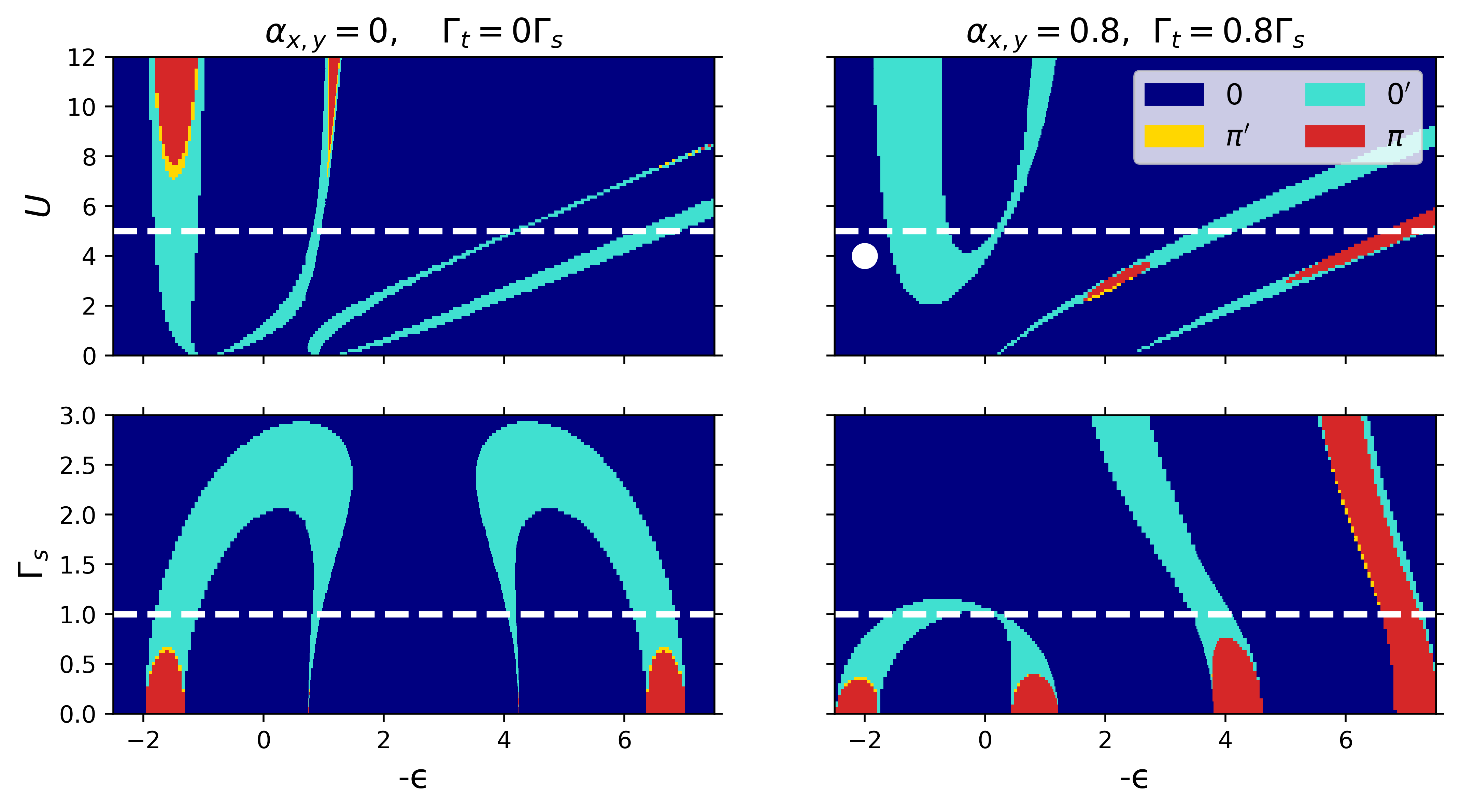}
    \caption{Phase diagrams of the four-sites model in the \(U{-}\epsilon\) and \(\Gamma_s{-}\epsilon\) planes, with and without SO respectively in the right and left columns. The white dashed lines in the upper (lower) row correspond to the values of \(U\) (\(\Gamma_s\)) used in the lower (upper) row. The white point in the upper right figure corresponds to the parameters used for Fig. 3 in the main text (hopping parameters \(t_{x,y}\) are fixed to \({-}1\)).}
    \label{fig:phase_diagram}
\end{figure}

To get an idea of the main properties of this model we show in Fig. \ref{fig:phase_diagram} some phase diagrams for different parameter choices.  
If the normal region of the nanowire could become more isolated from the leads, we would expect a QD-like behavior. 
The most typical feature, that arises from the interplay of the superconducting pairing, the Coulomb interaction and the coupling with the leads, is the transition to a \(\pi\)-junction behaviour where the ground state (GS) changes parity. 
In Fig. \ref{fig:phase_diagram} the colors indicate the phases ``\(0\)'', where the GS is even \(\forall\delta\) and the absolute minimum is at \(\delta{=}0\) (dark blue); ``\(\pi\)'', where the GS is odd \(\forall\delta\) and the absolute minimum is at \(\delta{=}\pi\) (red); and ``\(0'\)'' and ``\(\pi'\)'' (bright blue and yellow), which are intermediate phases similar to the previous ones, but where the parity of the GS is not the same \(\forall\delta\). The top row shows diagrams in the \(U{-}\epsilon\) plane, where a ``\(0\)'' background develops vertical and diagonal regions with different phase at sufficiently high value of \(U\). Their structure is similar to the diagram associated to linear arrays of quantum dots between superconducting leads \cite{QDarray} when the number of dots is 4. As discussed in that reference, for a sufficiently large fixed interaction and weak coupling to the leads, the GS alternates parity as the dots filling increases (i.e. for increasing \({-}\epsilon\)).
The figures in the bottom row are diagrams in the \(\Gamma_s{-}\epsilon\) plane, displaying $0'$ regions with inverted ``U'' shapes that connect odd valleys. As can be observed in the right lower panel these regions become distorted when spin-orbit interactions is switched on.

\section{S4. Evolution of the transition lines}
\label{sec:S4}

The rich structure of the transition lines, characterized by the 4-fold degeneracies at phases \(\delta{=}0,\pi\) of the odd transitions and the full splitting of the even ones, emerges from the presence of time reversal symmetry and the combination of spin-orbit coupling with Coulomb interaction. In this section, we progressively describe how these ingredients affect the ABSs energy spectrum, with the consequent repercussion in the evolution of the transition lines.

\begin{figure}[b]
    \centering
    \includegraphics[width=1\linewidth]{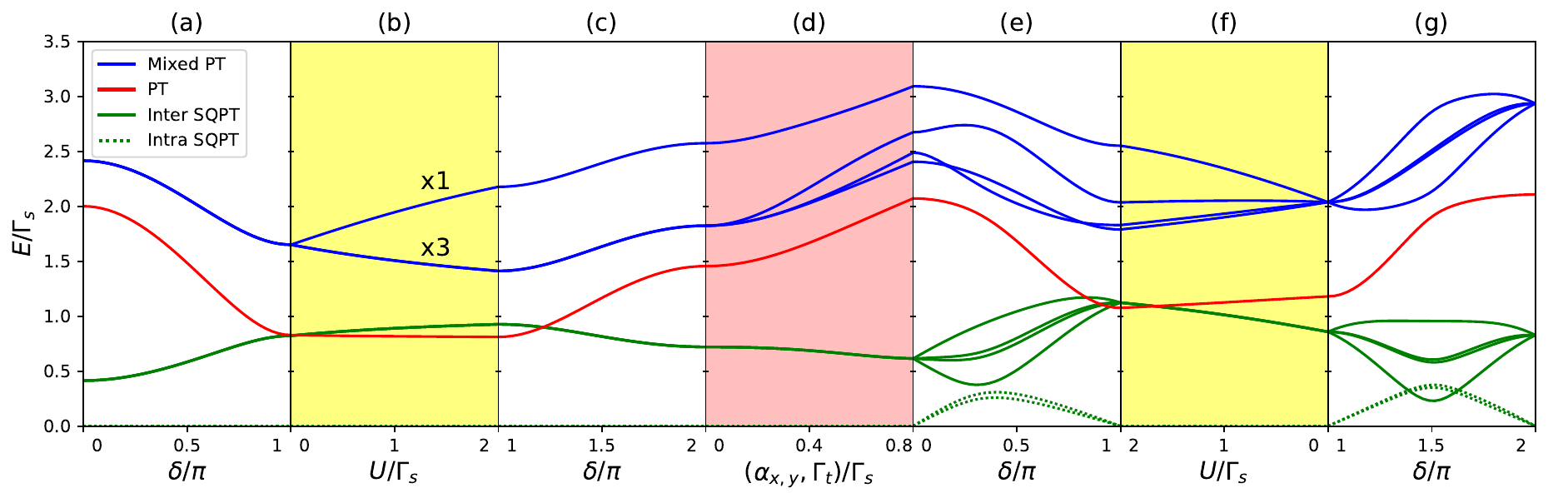}
    \caption{Evolution of the transitions in the 4-sites model with phase difference (white background), SO (pink) and interaction (yellow). From left to right, it starts displaying the evolution in phase difference \(\delta\) without SO nor interaction (a), then it includes interactions at \(\delta{=}\pi\) (b), and evolves again in \(\delta\) (c). In (d), it includes SO at \(\delta{=}0\), then evolves in \(\delta\) (e) and starts removing the interaction at \(\delta{=}\pi\) (f). Finally, in (g) it shows the evolution in \(\delta\) with SO but without interaction. Fixed parameters are \(\epsilon_{i,\tau}{=}1.5\Gamma_s\), \(t_x{=}2t_y{=}{-}\Gamma_s\). The higher PT to the second manifold is not shown because it can not be distinguished from other mixed PTs involving higher manifolds.}
    \label{fig:evo}
\end{figure}

In the situation without spin-orbit nor Coulomb interaction, the four ABSs of lowest energy, which correspond to the odd states with 1 QP, consist of two spin degenerated manifolds \(\forall \delta\). In the even sector, there are 6 states made of 2 QPs: 2 states where both QPs are in the same manifold with opposite spin (they give rise to pair transitions from the ground state), and 4 degenerate states where each QP is in one different manifold (these give rise to mixed pair transitions) (Fig. \ref{fig:evo}a). When the interaction is introduced, the odd states maintain their degeneracies, and the mixed even states split in a singlet plus a triplet. This behaviour stems from the spin rotational symmetry, encoded in \([H_0{+}H_{int}, S_i]{=}0\), since for any state with certain energy and spin, there is another state with the same energy but with rotated spin (same total spin, different spin projection) (Fig. \ref{fig:evo}b,c).

In the non-interacting situation with spin-orbit, spin is no longer a good quantum number. This allows for a splitting in almost all \(\delta\)'s, which is obtained for long multichannel junctions. However, time reversal symmetry imposes some constraints. Firstly, since the phase difference is \(2\pi\)-periodic and ultimately originates from a magnetic flux, we have, respectively, \(H(\delta{+2\pi}){=}H(\delta)\) and \(\mathcal{T}H(\delta)\mathcal{T}^{-1}{=}H(-\delta)\), so the spectrum over \(\delta\) must be mirror-symmetric around \(\delta{=}0,\pi\) (this constraint also applied for the previous situation without SO). Secondly, since in the odd states there is always at least one unpaired spin and \(\mathcal{T}\) reverses it, there must be pairs of odd states with the same energy (Kramers degeneracy) (Fig. \ref{fig:evo}d-g). Mixed even states inherit this degeneracy when no interactions are present, but when they are, no constraints prevent the splitting (Fig. \ref{fig:evo}d-f).

However, it must be noticed that if the nanowire is symmetric in the transverse direction (\(\epsilon_{i,\tau}{=}\epsilon_{i,\bar{\tau}}\)), the hybridization of the spin with the translational degrees of freedom produced by the Rashba SO still conserves a \textit{mirror} symmetry that composes the spin in the transverse direction (\(y\)) with a spatial mirroring over it \cite{Hays2021}. This symmetry can be described with the operator \(M_y{=}\sum_{i,\tau,\sigma,\sigma'} c^{\dagger}_{i,\bar{\tau},\sigma}(\sigma_{y})_{\sigma,\sigma'}c_{i,\tau,\sigma'}\), which, in a mirror symmetric situation (\([H_0, M_y]{=}0\)), has odd (even) integer eigenvalues for odd (even) states. The presence of this symmetry can lead us to think on the possibility of a similar argument as that of the Kramers theorem protecting some of the degeneracies in the mixed even states, since e.g. the mixed states depicted in Fig. \ref{fig:evo}g have pseudospin \(m_y{=}0,0,-2,2\), and the state with \(m_y{=}2\) at certain \(\delta\) has to be the time reversed of the state with \(m_y{=}-2\) at \(-\delta\), with the same energy. However, quite surprisingly, the onsite interaction does not conserve the pseudospin symmetry (\([H_{int}, M_y]{\neq}0\)), so the argument does not hold and the degeneracy might be broken.

\section{S5. Experimental details}
\label{sec:S5}

\begin{figure*}[h]
    \centering
    \includegraphics[width=0.6\textwidth]{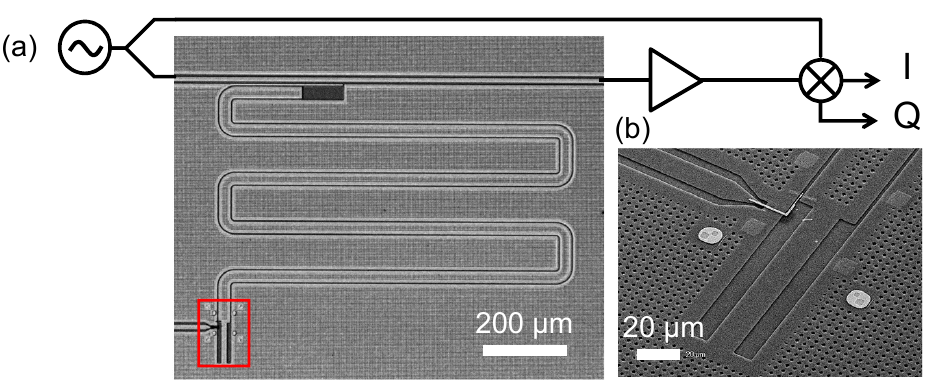}
    \caption{Optical image: Large scale view of the hanger-style microwave resonator, coupled to a measurement bus line. The readout tone is sent through the bus line, then amplified (triangle) with a TWPA followed by a HEMT and a room-temperature amplifier; and finally down-converted with an IQ mixer. A superconducting coil placed under the sample allows to control the superconducting phase difference across the weak link.}
    \label{fig:resonator}
\end{figure*}

The InAs/Al nanowire used in this experiment is from the same batch as in Ref.~\cite{Tosi2019}, with an InAs core diameter of 140~nm, and a 25~nm thick Al full shell. Figure~\ref{fig:resonator} shows a mm-scale view of the hanger-style CPW quarter-wavelength resonator and a tilted closer view of its shorted end. The nanowire bridges between the central line of the resonator and one of the side ground planes, to form a $1000~\mu$m$^2$ superconducting loop used to phase-bias the weak link. A flux quantum through the loop corresponds to an applied magnetic field of $\sim$2~µT.  This new design, in which the loop is part of the resonator itself, results in a stronger weak link-resonator coupling than in Ref.~\cite{Tosi2019}. In addition, placing the gate directly under the weak link instead of on its side improves the spectra stability, probably because the gate screens the charge fluctuations at the substrate surface. Altogether, these changes lead to higher quality spectra, as shown already in Ref.~\cite{Metzger2021}.  Figure~\ref{fig:wiring} shows the wiring of the experiment within the dry dilution fridge in which it was measured (base temperature: 30~mK).

\begin{figure*}[h]
    \centering
    \includegraphics[width=\textwidth]{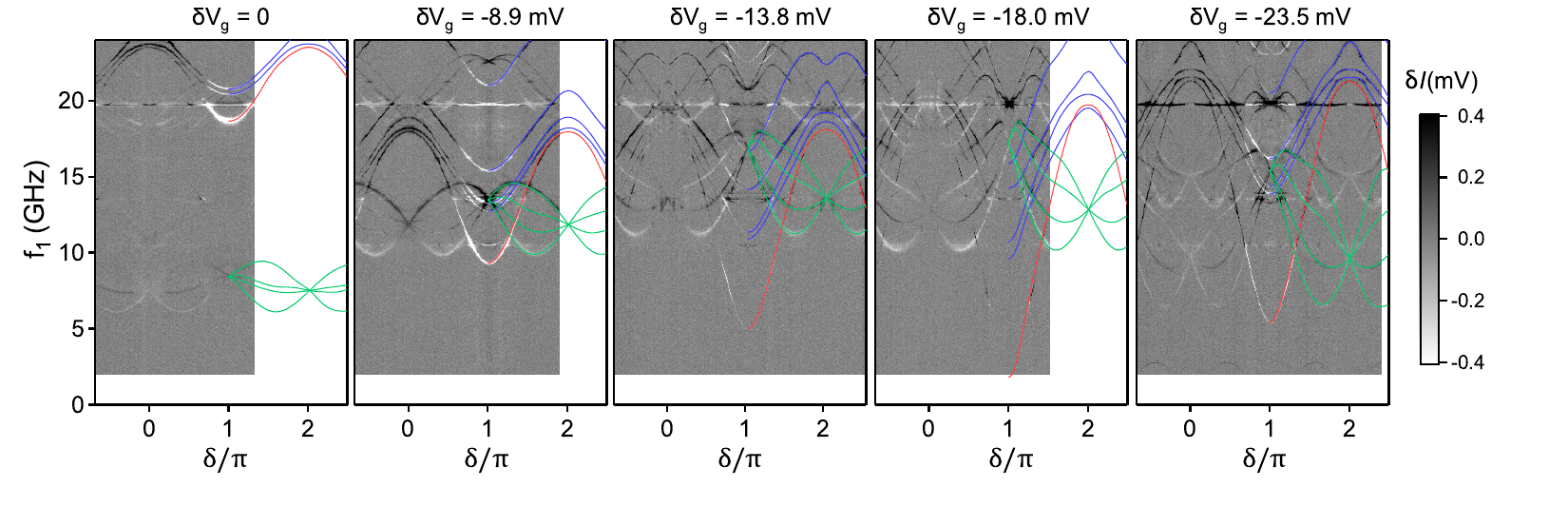}
    \caption{In complement to Fig.~2: data on the other quadrature ($\delta I$). We show here the data in the interval $\delta=-0.68\pi$ to $2.5\pi,$ without completing with $2\pi$-shifted copies as was done in Fig.~2.}
    \label{fig:dataBis}
\end{figure*}

All data in Fig.~\ref{Fig:Fig2} were not taken on the interval shown: the figure was completed by $2\pi$-shifted copies of the measured data. In Fig.~\ref{fig:dataBis}, which displays  the other quadrature, we only show the data as taken between $\delta=-0.68\pi$ and $\delta=2.5\pi.$

We note that intra-manifold transitions are not seen in these spectra. As a matter of fact, those transitions lines are faint \cite{Metzger2021}, and hardly visible at moderate drive power.

\begin{figure*}[h]
    \centering
    \includegraphics[width=\textwidth]{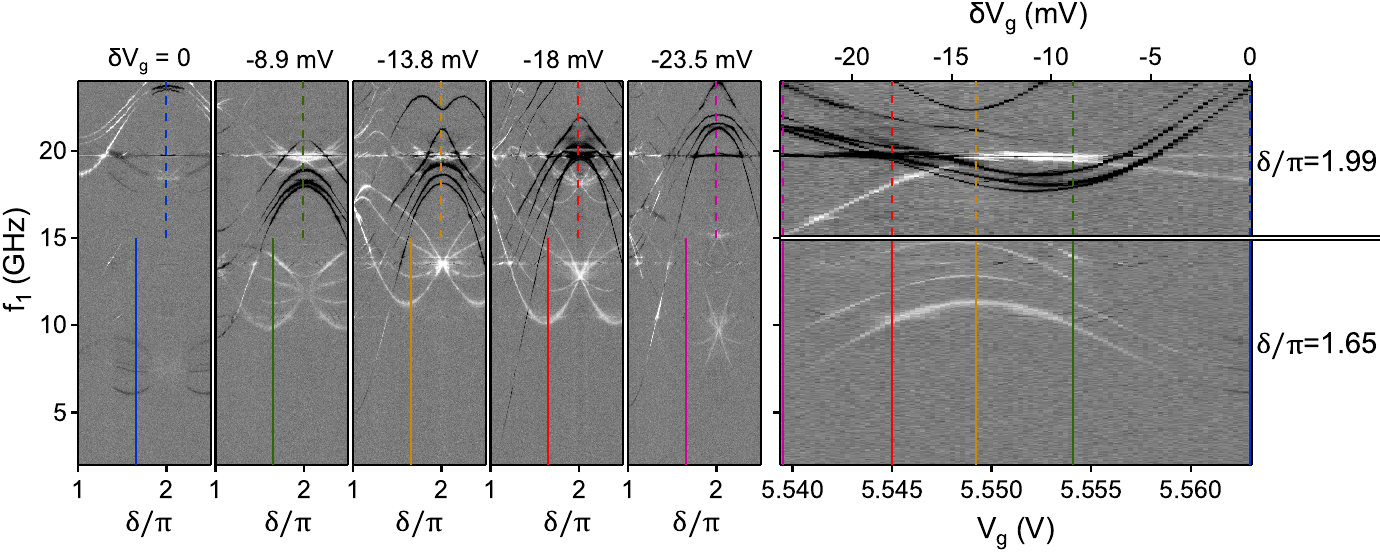}
    \caption{Rightmost panel: Evolution of the spectrum with gate voltage (data on $Q$ quadrature, same grey scale as Fig.~2), at $\delta=1.65\pi$ from $2$ to $15$~GHz, and at $\delta=1.99\pi$ from $15$ to $24$~GHz. Solid and dashed vertical color line correspond to the gate voltages at which the spectra of Fig.~2 were measured. In the other panels, we reproduce the right half of the spectra of Fig.~2, with lines that mark the values of $\delta$ at which the gate sweep was taken, allowing to make a correspondence between the panels.}
    \label{fig:gateSweep}
\end{figure*}

Figure~\ref{fig:gateSweep} shows, in the rightmost panel, the evolution of the spectrum as a function of the gate voltage at values corresponding to Fig.~2. For frequencies below $15$~GHz, the phase was $\delta=1.65\pi,$ whereas the top part, above $15$~GHz, was taken at $\delta=1.99\pi.$  Note that the panels showing the spectra as a function of phase, repeated here for ease of correspondence with the gate evolution, are ordered with decreasing gate voltage from left to right, whereas the direction of the gate voltage axis in the rightmost panel is opposite. The evolution is smooth, without any jump.

\begin{figure*}[h]
    \centering
    \includegraphics[width=\textwidth]{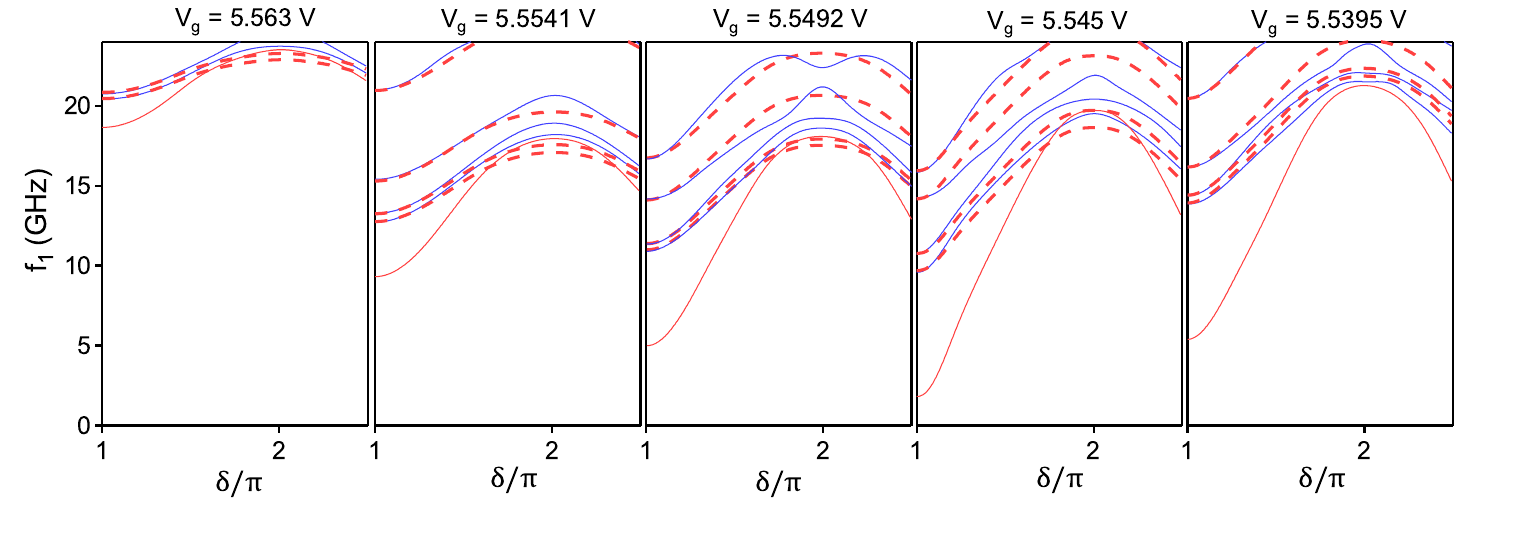}
    \caption{Red and blue lines are the splines overlying the data in Fig.~2 and Fig.~\ref{fig:dataBis}. Dashed red lines are obtained by taking half the frequency of the red lines, and shifting vertically. This shows that, around $\delta=\pi,$ the dispersion of the mixed pair transition have a curvature close to half that of the lowest pair transition.}
    \label{fig:mysteryLines}
\end{figure*}

\begin{figure*}[h]
    \centering
    \includegraphics[width=\textwidth]{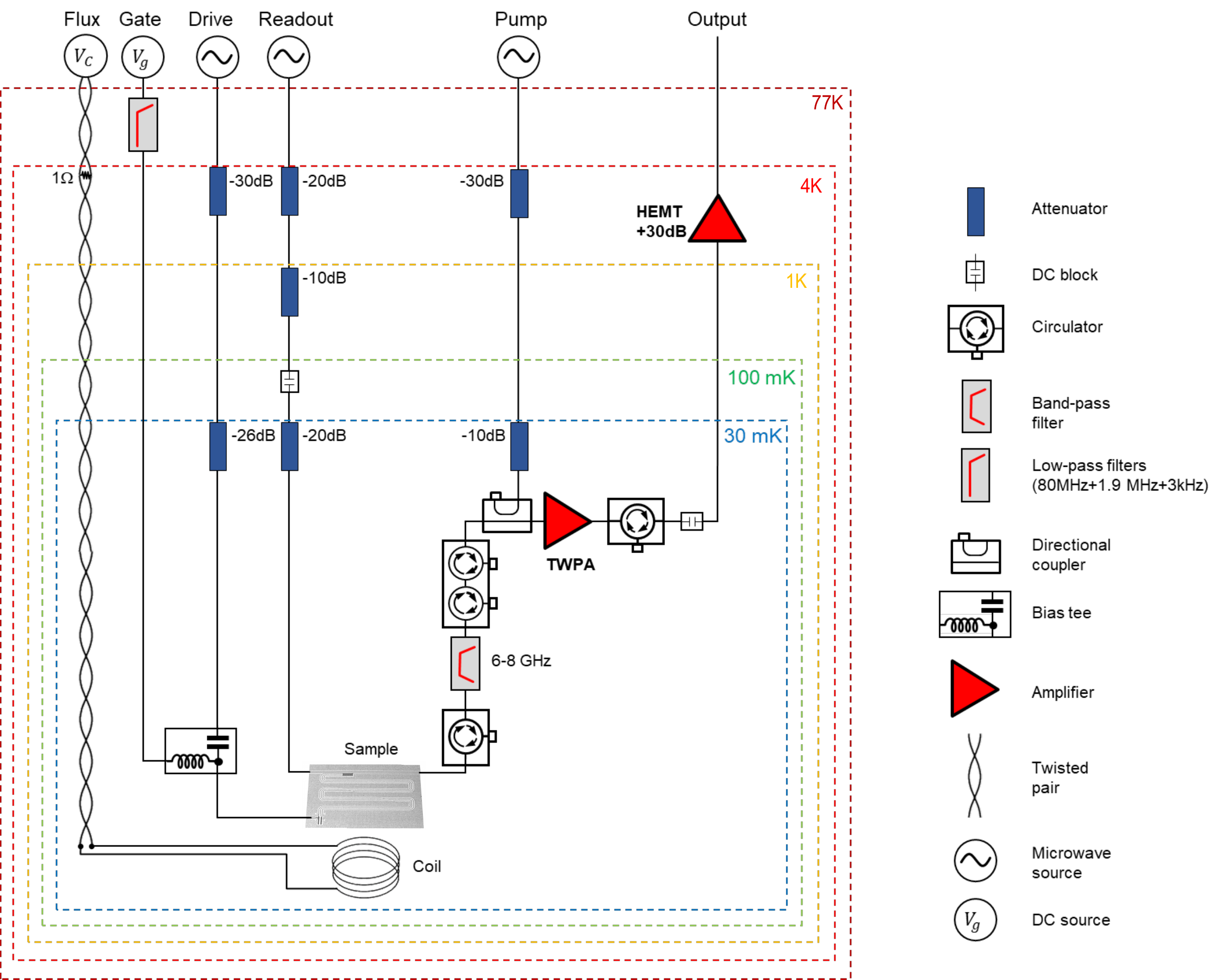}
    \caption{Cryogenic wiring of microwave and DC connections to the sample. A superconducting coil placed under the sample allows to control the superconducting phase difference across the weak link. The measurement signal is amplified by a TWPA at base temperature, followed by a HEMT at the 4K stage.}
    \label{fig:wiring}
\end{figure*}

As a final remark, we note that when the first manifold has a stronger curvature than the second one, the dispersion of mixed pair transition lines near $\delta=\pi$ resembles half that of the pair transition to the lowest manifold, plus an offset due to the energy of the second manifold and to interactions. This is illustrated in Fig.~\ref{fig:mysteryLines}. This might be an explanation for the ``mystery lines'' discussed in Ref.~\cite{Hays2020} and for the additional transition lines discussed in the SM of Ref.~\cite{Fatemi2021}, where the presence of non-dispersing, localized states was invoked.

\clearfmfn
\newpage

\end{document}